\documentclass[12pt,preprint]{aastex}

\shorttitle{}
\shortauthors{}

\usepackage{epsfig}
\usepackage{multirow}
\usepackage{multicol}
\usepackage{psfig}
\usepackage{color}
\usepackage{amsmath}
\usepackage{amssymb}
\usepackage{extsizes}
\usepackage{subfigure}
\usepackage{verbatim}
\usepackage{rotating}

\newcommand{\be}{\begin{eqnarray}}
\newcommand{\ee}{\end{eqnarray}}
\newcommand{\ben}{\begin{enumerate}}
\newcommand{\een}{\end{enumerate}}

\newcommand{\bmc}{\begin{multicols}{2}}
\newcommand{\emc}{\end{multicols}{2}}

\newcommand{\btable}{\begin{table}[htb]  \begin{center}  \begin{tabular}}
\newcommand{\etable}{\end{tabular}  \end{center}    \end{table}}

\newcommand{\ns}{{\mbox N_{\textrm{side}}}}

\def\leqsim{\mathrel{\lower2.3pt\hbox{$\mathpalette\oversim<$}}}
\def\geqsim{\mathrel{\lower2.3pt\hbox{$\mathpalette\oversim>$}}}
\def\oversim#1#2{\lower.2pt\vbox{\baselineskip0pt\lineskip-.5pt
  \ialign{$\mathsurround=0pt #1\hfill##\hfil$\crcr#2\crcr\sim\crcr}}}

\begin{document}

\title{Excess ellipticity of hot and cold spots in the WMAP data?}

\author{Eirik Berntsen}
\affil{Institute of Theoretical Astrophysics, University of Oslo,
P.O. Box 1029 Blindern, N-0315 Oslo, Norway}
\email{baardeb@astro.uio.no}

\author{Frode K. Hansen}
\affil{Institute of Theoretical Astrophysics, University of Oslo,
P.O. Box 1029 Blindern, N-0315 Oslo, Norway} 
\email{frodekh@astro.uio.no}

\begin{abstract}
We investigate claims of excess ellipticity of hot and cold spots in the WMAP data (Gurzadyan et al. 2005, 2007).
Using the cosmic microwave background data from 7 years of observations by the WMAP satellite, we find, contrary to previous claims of a $10\sigma$ detection of excess ellipticity in the 3-year data, that the ellipticity of hot and cold spots are perfectly consistent with simulated CMB maps based on the concordance cosmology. We further test for excess obliquity and excess skewness/kurtosis of ellipticity and obliquity and find the WMAP7 data consistent with Gaussian simulated maps.
\end{abstract}

\keywords{ (cosmology:) cosmic microwave background --- cosmology:
  observations --- methods: data analysis ---  methods: statistical
  --- ellipticity}

\section{Introduction}
\label{sect:intro}

Ever since the discovery of the cosmic microwave background (CMB) fluctuations in
1992 \citep{smoot} the availability and quality of CMB data
have been steadily increasing through a series of experiments. The
{\emph{Wilkinson Microwave Anisotropy Probe (WMAP)}
  \citep{Bennett:2003bz} with its most recent seven year maps
  \citep{Hinshaw:2008kr} is currently the most important publicly available data set for
  the study of CMB fluctuations. In the standard model of cosmology
the universe is asymptotically isotropic and homogeneous on large
scales and the CMB consists of Gaussian
fluctuations making the fluctuation \emph{amplitude} on different
scales the only relevant information contained in the map. 

Studies
of CMB fluctuations have classically concentrated on this; achieved most commonly by
the transformation of the map to spherical harmonic space
and averaging over the directional modes
to create the angular power spectrum. The power spectrum, while
computationally more forgiving to work with than the original map, is
only free of degeneracies under the assumption of perfect Gaussianity and
isotropy. 

In this paper we will study a property of CMB fluctuations directly related to the map,
namely the ellipticity of hot and cold spots. The aim of this study is to follow up on direct
measures of ellipticity of spots in CMB data in \cite{Gurzadyan:2002cw,
  Gurzadyan:2003xh, Gurzadyan:2004mn, Gurzadyan:2005uw,
  Gurzadyan:2006uk} where a significant deviation from the ellipticity expected 
for Gaussian fluctuations was found.  It has been shown that within
curvature differences of $\Delta \Omega_{\textrm{tot}} = 0.05$,
  standard cosmological models produce undetectably small effects
  in CMB ellipticity  compared to a perfectly flat universe
  \citep{Aurich}, so that any excess ellipticity found in \emph{WMAP}
  would be a significant indication of non-standard physics. For a more complete discussion, see \cite{Aurich}.

 We perform our own analysis of ellipticity on the
\emph{WMAP} 7-year data and compare to results stated in the given
references, in particular we wish to test the claims of significantly
higher ellipticity in \emph{WMAP} data than in simulations
and the claims for higher ellipticity on smaller scales. It is also
claimed that there is no evidence of a preferred direction of the spot elongations 
and we investigate this by obliquity measures of
the CMB data.

In \S \ref{sect:data} we describe the data and masks used in the
analysis. The methods used to assess the ellipticity and obliquity are
outlined in \S \ref{sect:method}. In \S \ref{sect:res} we show the
results of the ellipticity and obliquity tests applied to the 7-year
WMAP data and in
\S \ref{sect:conclusions} we conclude.

\section{Data}
\label{sect:data}

 The analysis in this paper was performed using the seven year release of
 the WMAP data (publicly available at the Lambda web
 site\footnote{http://lambda.gsfc.nasa.gov/}) as well as a statistical
 ensemble of 5000 simulated maps of each of the channels Q (41GHz), V
 (61GHz) and W (94GHz) (the map for each channel is obtained by
 taking the mean of all differencing assemblies
 for the given channel). All analysis are preformed on either the Q band or the co-added V+W
 band maps.

 From
 all maps, we have subtracted the best fit mono- and dipole. The mask
 used throughout was the WMAP KQ85 galactic and point source mask leaving a sky
 fraction of 82\%. Maps were simulated using WMAP noise and beam properties. 
 All maps were pixelized in the HEALPix
 scheme\footnote{http://healpix.jpl.nasa.gov/} \citep{healpix}

\section{Methodology}
\label{sect:method}
As described in \cite{Gurzadyan:2002cw} the method of extracting
anisotropic areas is to consider as relevant only those pixels which
temperature value is above (or below) a given temperature
threshold and as non relevant all pixels which temperature value is
below (above) the threshold.
A ``spot'' is then defined to be any set of relevant pixels such that
\begin{enumerate}
\item any pixel in the set may be path connected on the HEALPix map with any
other within the set without moving through a non relevant pixel
\item no pixel in the set may be path connected on the map with a relevant pixel
  outside the set without moving through a non relevant pixel.
\end{enumerate}
Connectedness is allowed on the diagonal, so that
two relevant pixels sharing only a vertex are still connected.
The ellipticity measures procedure differs slightly from
the one described in \cite{Gurzadyan:2002cw}.
Ellipticity for a spot as shown on \ref{fig:spot} is
measured directly on the sphere with the following algorithm:
\begin{enumerate}
\item Double the map $N_{\textrm{side}}$.
\item Determine the two pixels in the spot, $A_1$
    and $A_2$, with the greatest angular distance between them, half of this angular distance
    is the semi major axis $a$.
\item Determine the shortest angular distance to this axis for all pixels in the spot.
The two pixels (on either side) with the largest distance to the axis are referred to as
  $B_1$ and $B_2$ and their angular distance to the major axis as
  $b_1$ and $b_2$ respectively.
\item The minor axis $b$ is then defined as
\begin{equation}
b = b_1 + b_2
\end{equation}
\item Calculate ellipticity, defined as
\begin{equation}
\epsilon \equiv \frac{a}{b}.
\end{equation}
\item Calculate the obliquity $\varepsilon$, defined as the smallest
  angle between the great circle traced by the major axis and a chosen
  equator.\label{stepOb}
\end{enumerate}
The first step is a purely computational one made to avoid cases where
all pixels in a spot fall
on a line, causing a potentially near infinite ellipticity as all
pixels return close to zero distance to the major axis.  This was seen
as superior to the use of vertex positions as a means for less skewed
results for smaller spots. Note that
any of the three geodesics $a$, $b_1$ and $b_2$ may trace areas not
covered by the spot as the major axis does in figure
\ref{fig:spot}. Except for a test of obliquity measures against 12288
chosen equators to check for differences, obliquity was measured against
galactic equator. Step \ref{stepOb} differs from
\cite{Gurzadyan:2002cw}, as it only measures obliquity from $0 - 90$
degrees to avoid ambiguity on polar spots.
\begin{figure}[ptb]
  \centering
  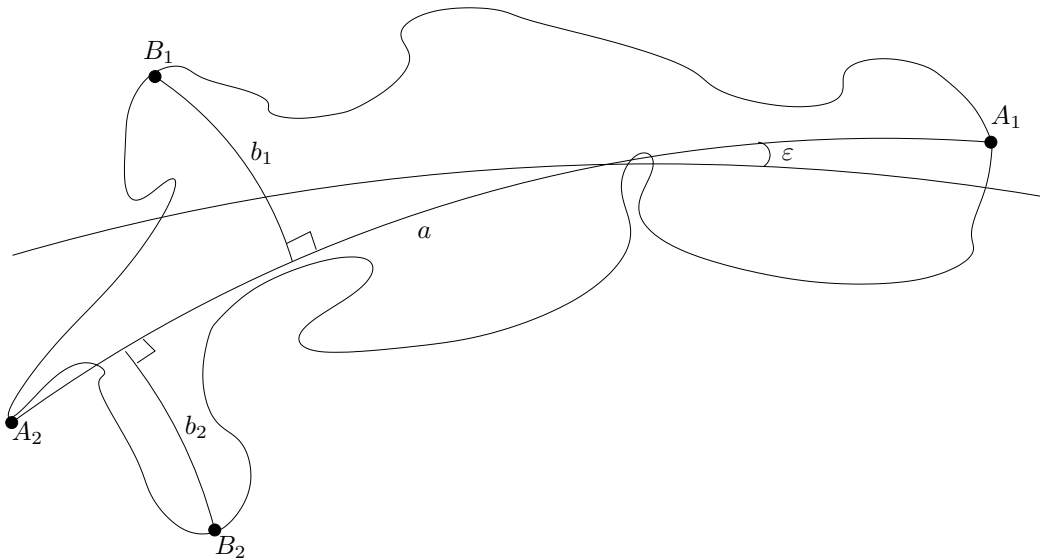
  \caption{Example of a spot (presumably with very high resolution). The major axis
    is formed between the two points furthest from each other, $A_1$
    and $A_2$. The smallest angle $\varepsilon$ to the equator (the
    line with no connecting points crossing the figure, so this is an
    equatorial spot) is the obliquity. The
    point $B_1$ is the furthest from the major axis on
    one side, the point $B_2$
    on the other side.}
  \label{fig:spot}
\end{figure}
Finally the results for one map is returned as the average over all spots and
statistical uncertainties with no weighting with regard to position or
size except that spots smaller than certain
values are excluded from some of the analysis. Statistical analysis
is carried out in frequentist manner to obtain mean, variance,
skewness and kurtosis% with
%distribution of values for ellipticity and obliquity all
%assumed to be Gaussian.

To obtain a result against which to compare the ellipticity and
obliquity values of the \emph{WMAP} maps; an ensemble of 5000 maps were
simulated using modules included in the HEALPix package.
The analysis was performed on simulated and \emph{WMAP} data in the Q,
and combined VW bands with thresholds
ranging from -500$\mu$K to -40$\mu$K for negatively defined spots
and then from positive 40$\mu$K to 500$\mu$K at
every 20th $\mu$K. The ellipticity and obliquity was calculated for
\begin{enumerate}
\item All spots (including the ones with only two pixels).
\item Spots with $>2$ pixels.
\item Spots with $>3$ pixels.
\item Spots with $>8$ pixels.
\item Spots with $>20$ pixels.
\item Spots with $>50$ pixels.
\item Spots with $>100$ pixels.
\item Spots with $>300$ pixels.
\end{enumerate}
Single pixel ``spots'' were ignored throughout.

In order to evaluate the significance of the results, we performed a $\chi^2$ test, 
\begin{equation}
\chi^2 = \sum_{tp}\sum_{t'p'}(x_{tp} - \langle x_{tp} \rangle)C^{-1}_{tp,t'p'}(x_{t'p'} - \langle x_{t'p'} \rangle)
\end{equation}
where $x_{tp}$ is the value (mean over all spots for that map) for the ellipticity or obliquity for the map to be tested at
a temperature threshold $t$ and for a pixel size $p$. $\langle x_{tp} \rangle$ is the mean over all simulated maps and $C_{tp,t'p'}$ is
the covariance matrix obtained from gaussian simulations based on the WMAP best fit power spectrum and noise model.

\section{Results}
\label{sect:res}
Results gathered for the V+W map is shown in figure \ref{fig:VWnumber}
(for number of spots) and figure \ref{fig:VWellip}
(ellipticity). We do not show plots for the Q band as
they are very similar to the VW results. The mean ellipticity value for
spots in a perfectly Gaussian map with infinite resolution
can be shown to be $\varepsilon \approx 1.648$ \citep{Aurich}, so this
is the naively expected result. Ellipticity significantly in excess of this number is only evident for
spots smaller than 8 pixels, for larger spots, the ellipticity
is close to the theoretical value. This is due to a higher probability for smaller spots of
consisting only of pixels placed vertically, horizontally or
diagonally on a row, resulting in such spots having ellipticity in
excess of 10, and hence skewing the results towards a higher mean
ellipticity. 

\begin{figure}[ptb]
  \centering
  \subfigure[$>3$ pixel spots]{\psfig{figure=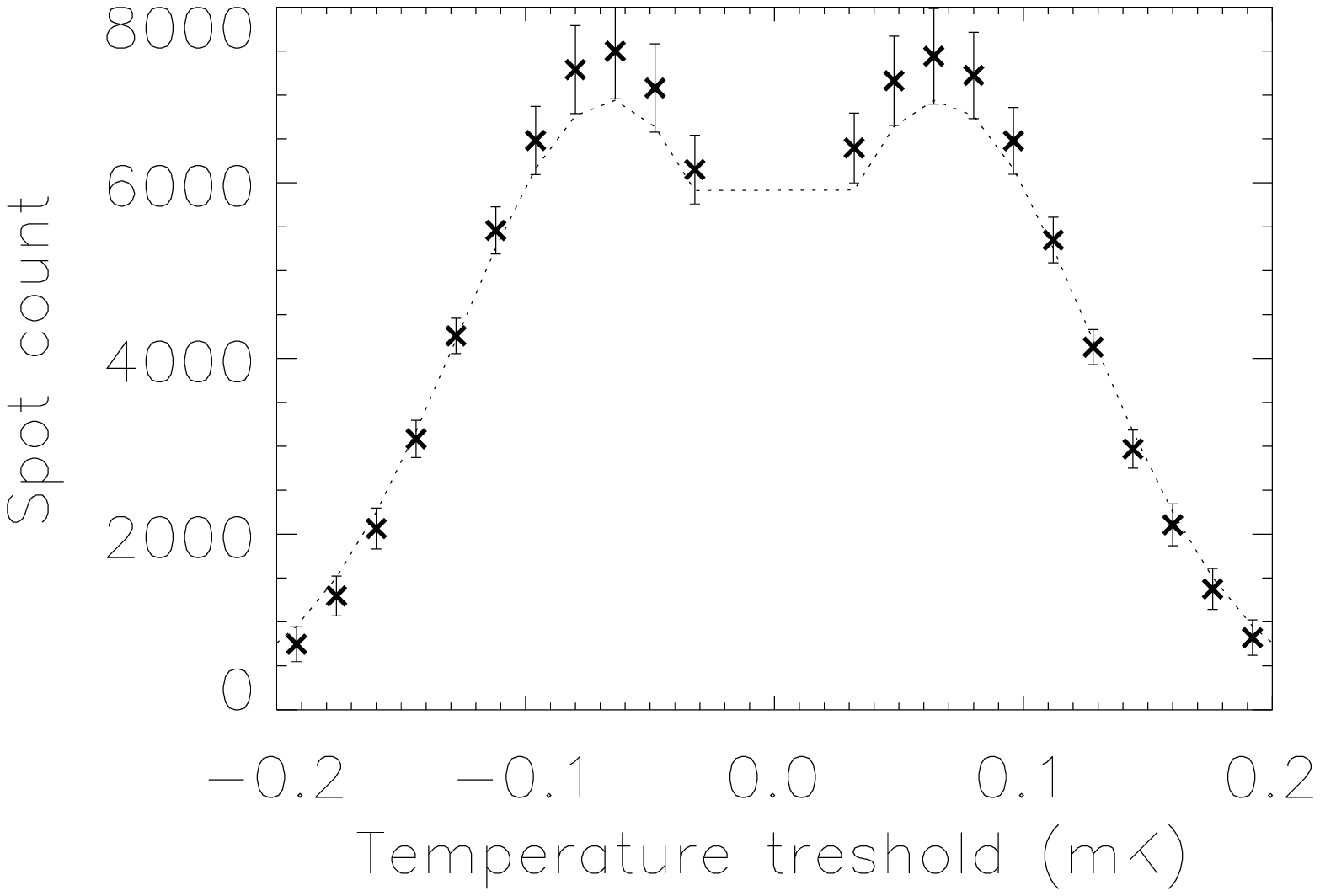,
      width=0.45\textwidth}}
  \subfigure[$>8$ pixel spots]{\psfig{figure=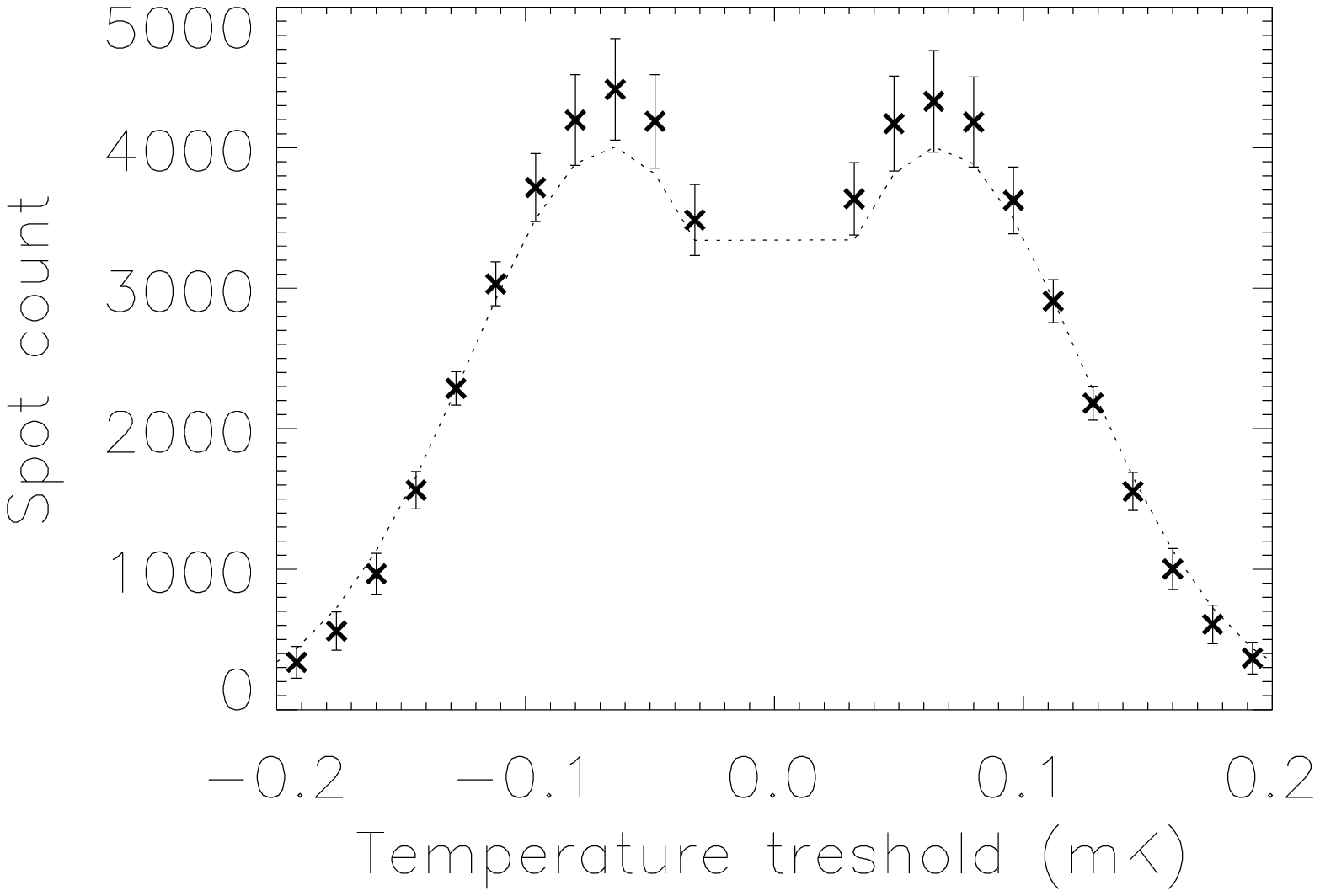,
      width=0.45\textwidth}}
  \subfigure[$>20$ pixel spots]{\psfig{figure=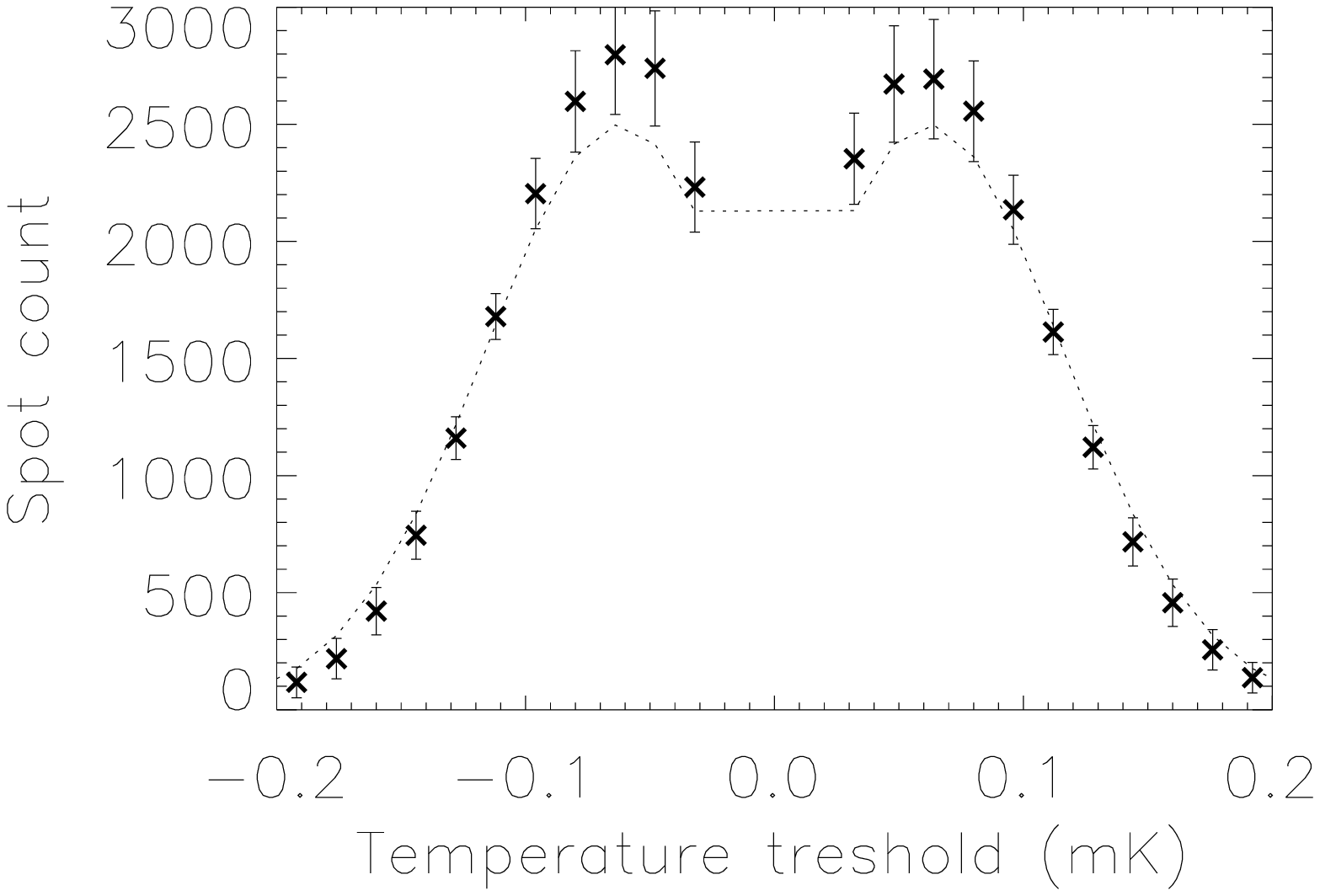,
      width=0.45\textwidth}}
  \subfigure[$>50$ pixel spots]{\psfig{figure=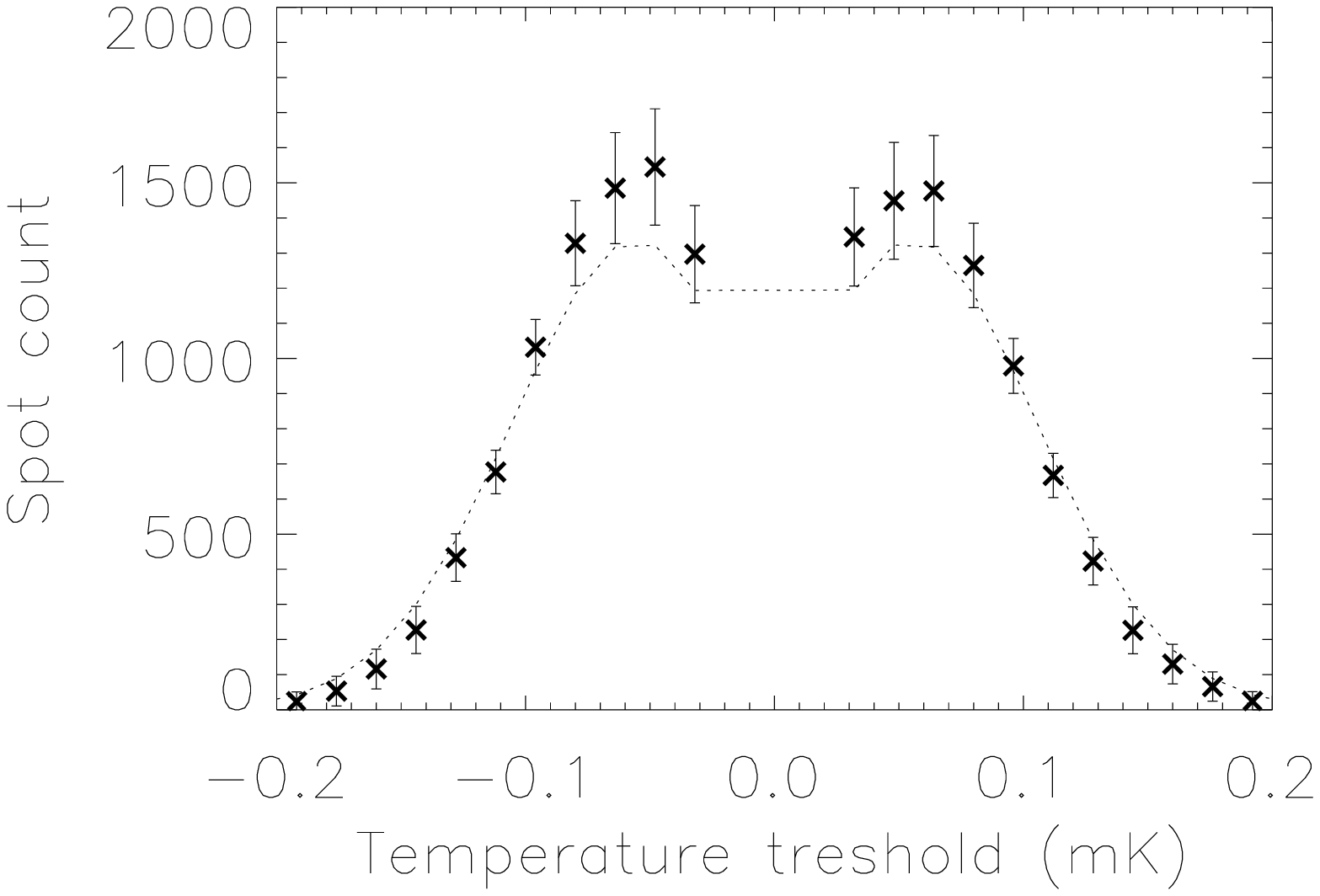,
      width=0.45\textwidth}}
  \subfigure[$>100$ pixel spots]{\psfig{figure=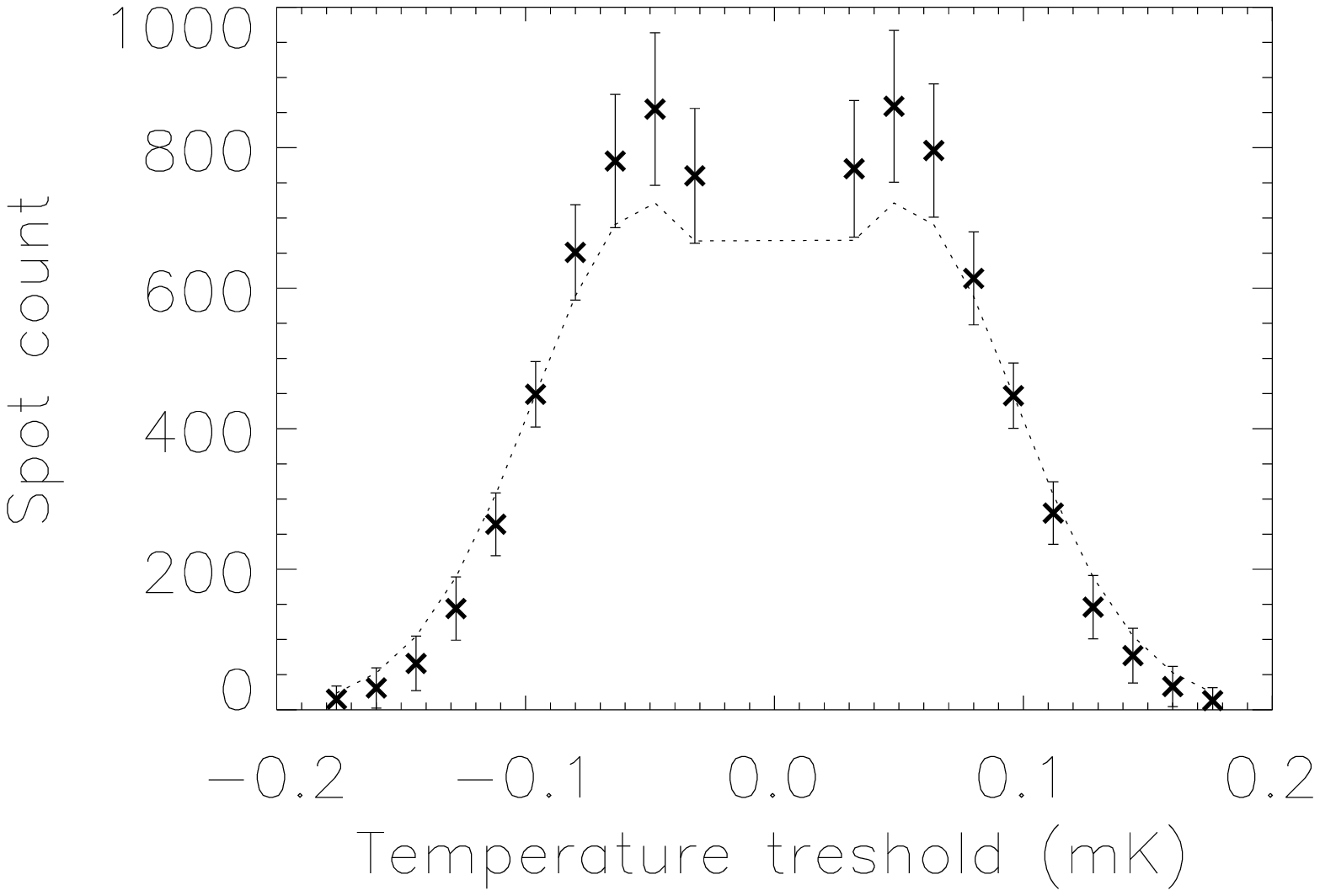,
      width=0.45\textwidth}}
  \subfigure[$>300$ pixel spots]{\psfig{figure=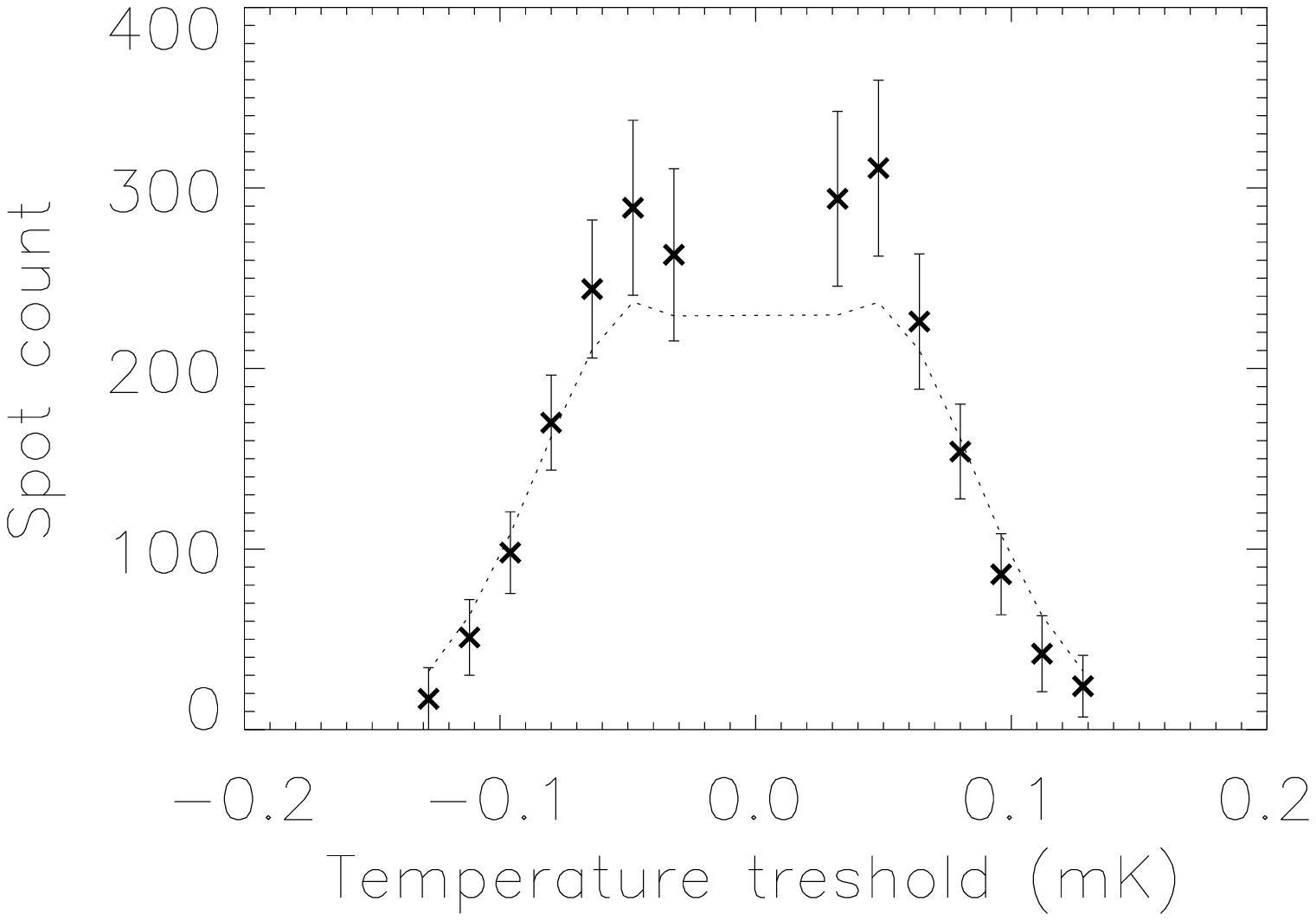,
      width=0.45\textwidth}}
  \caption{Number of spots in the V+W map as a function of threshold for a given size
  of spot. Data points are shown as crosses and the mean value from simulations as a dotted line. 
The confidence intervals are given to $2\sigma$. Note that we only show the points for which we had sufficient spot statistics to
be used in the analysis. Note also that the points are highly correlated.}
  \label{fig:VWnumber}
\end{figure}

\begin{figure}[ptb]
  \centering
  \subfigure[$>3$ pixel spots]{\psfig{figure=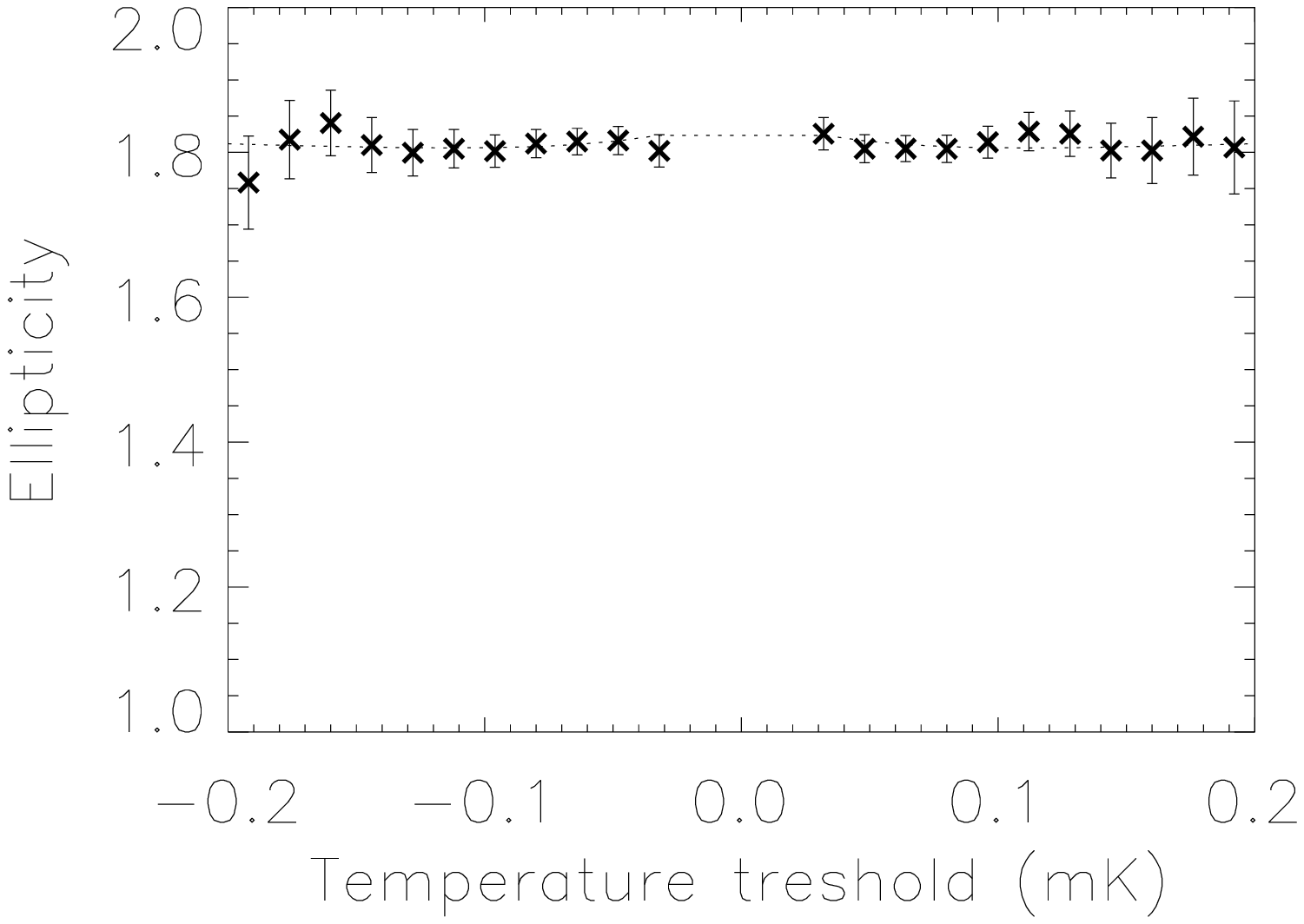,
      width=0.45\textwidth}}
  \subfigure[$>8$ pixel spots]{\psfig{figure=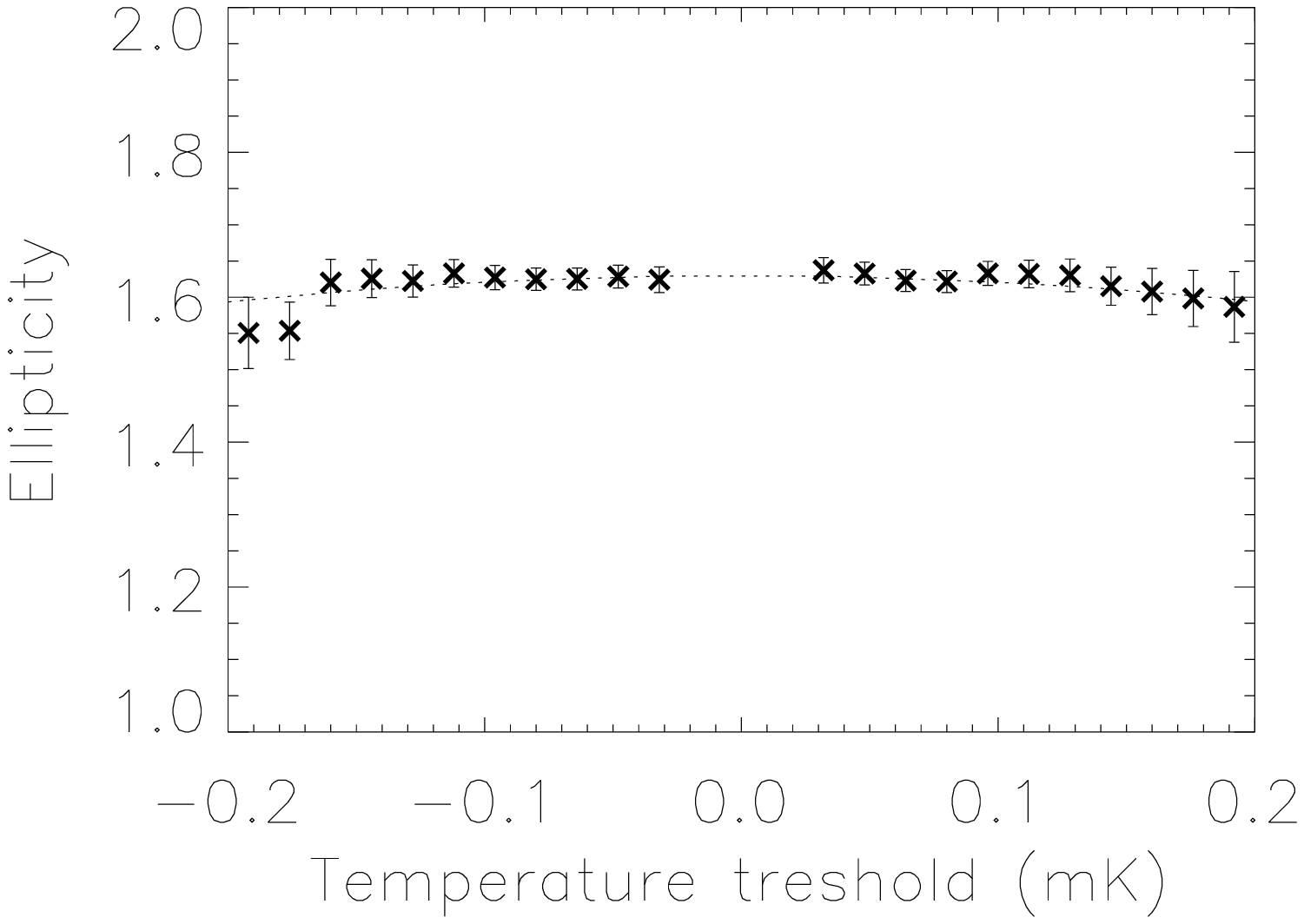,
      width=0.45\textwidth}}
  \subfigure[$>20$ pixel spots]{\psfig{figure=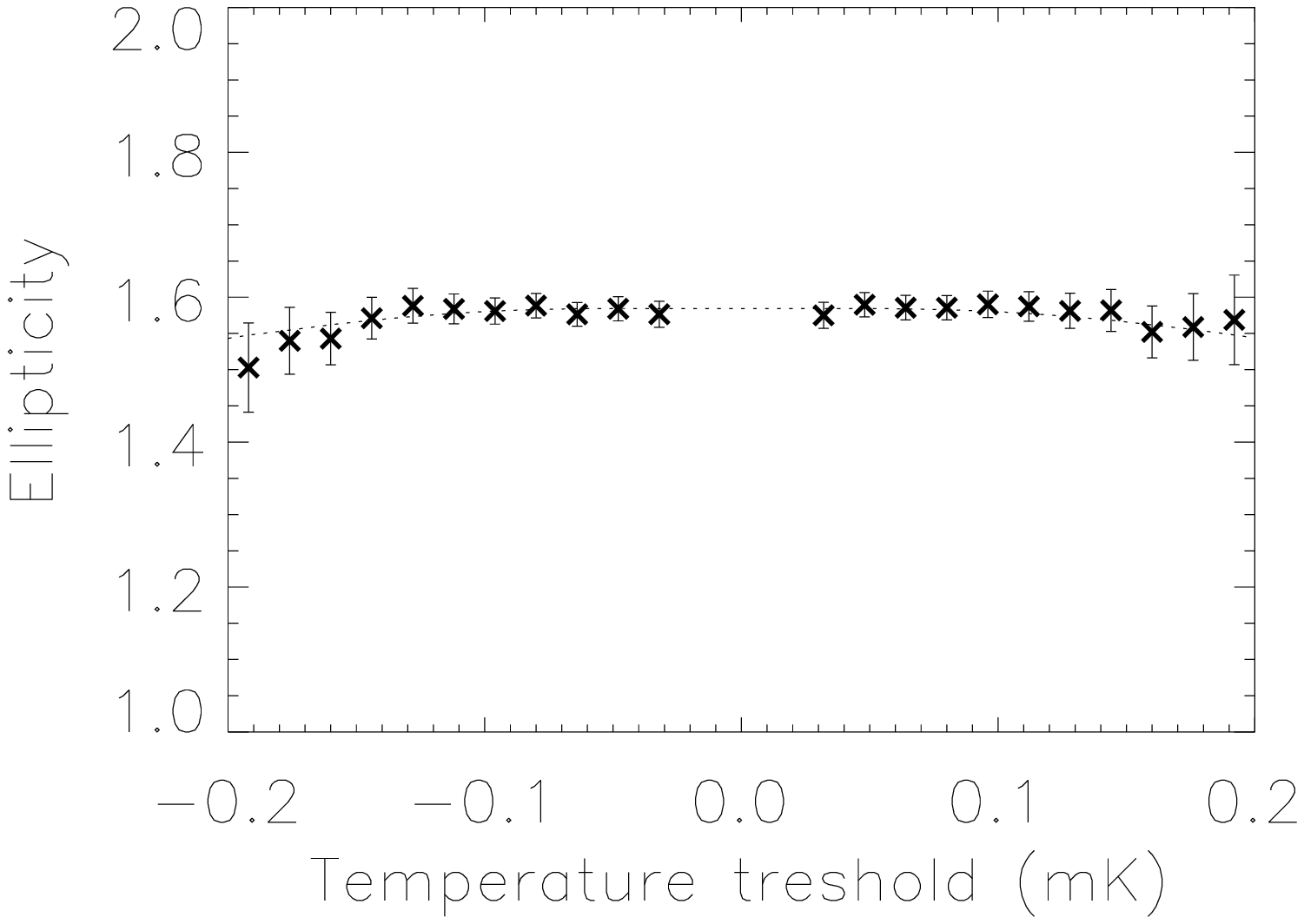,
      width=0.45\textwidth}}
  \subfigure[$>50$ pixel spots]{\psfig{figure=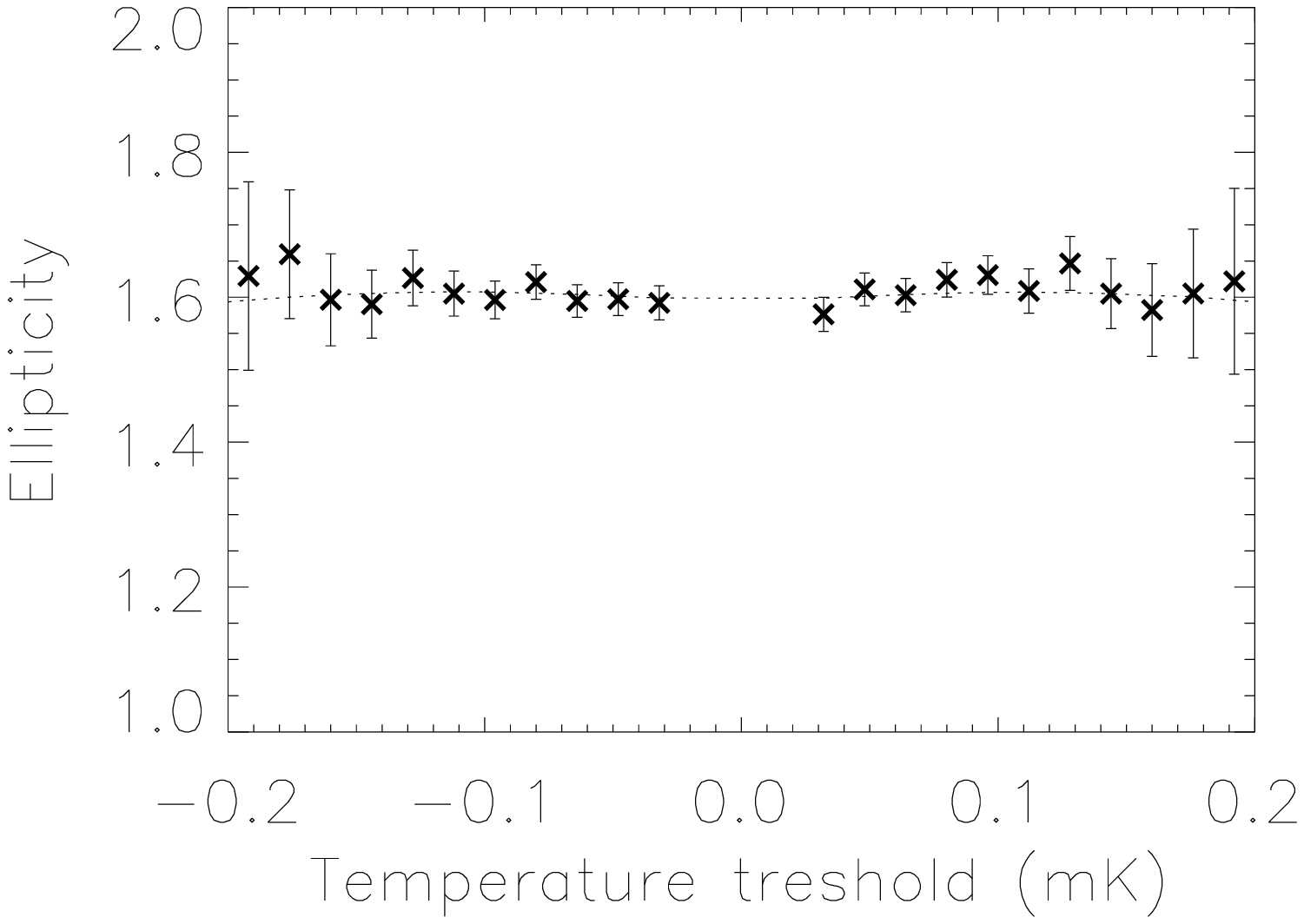,
      width=0.45\textwidth}}
  \subfigure[$>100$ pixel spots]{\psfig{figure=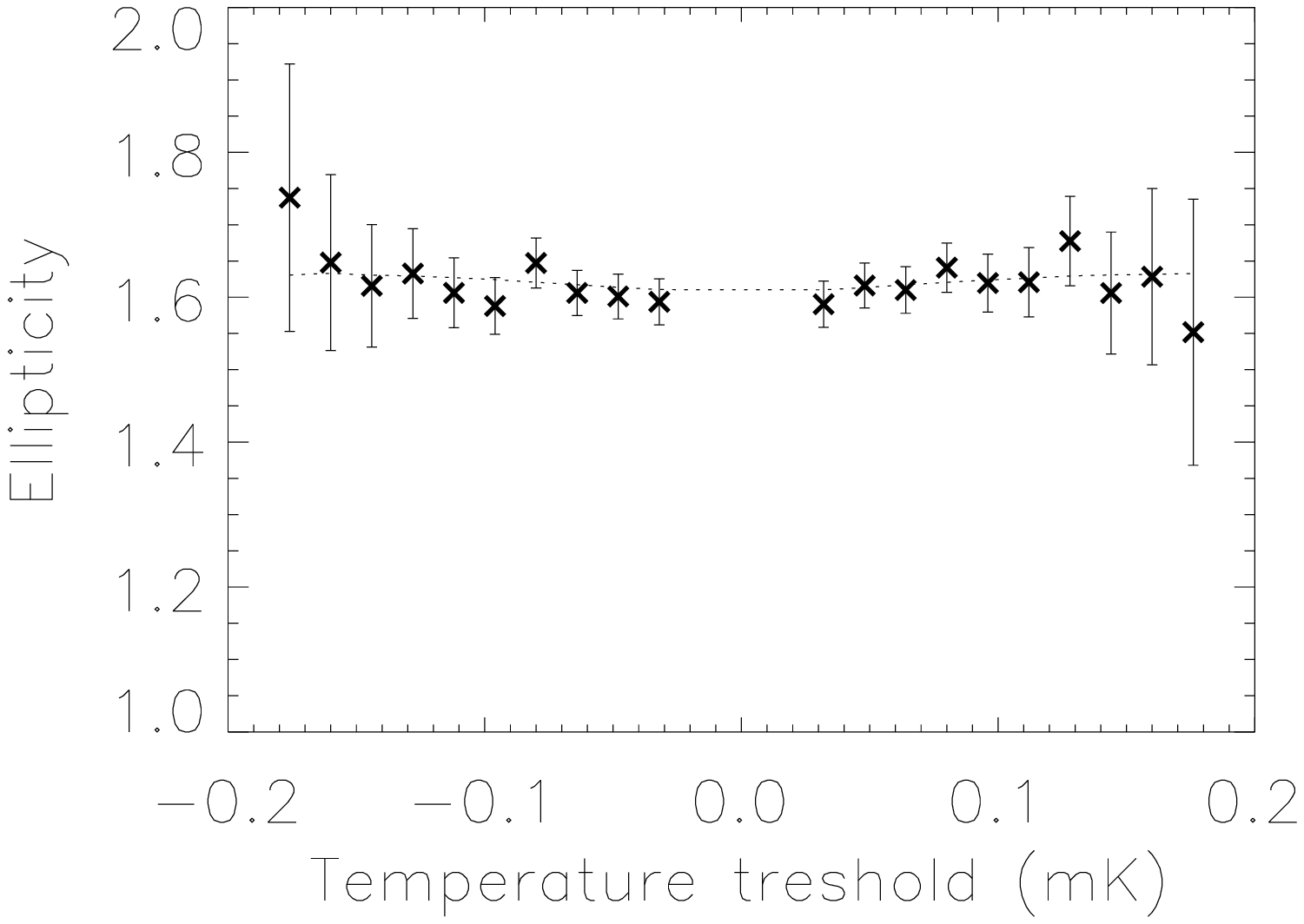,
      width=0.45\textwidth}}
  \subfigure[$>300$ pixel spots]{\psfig{figure=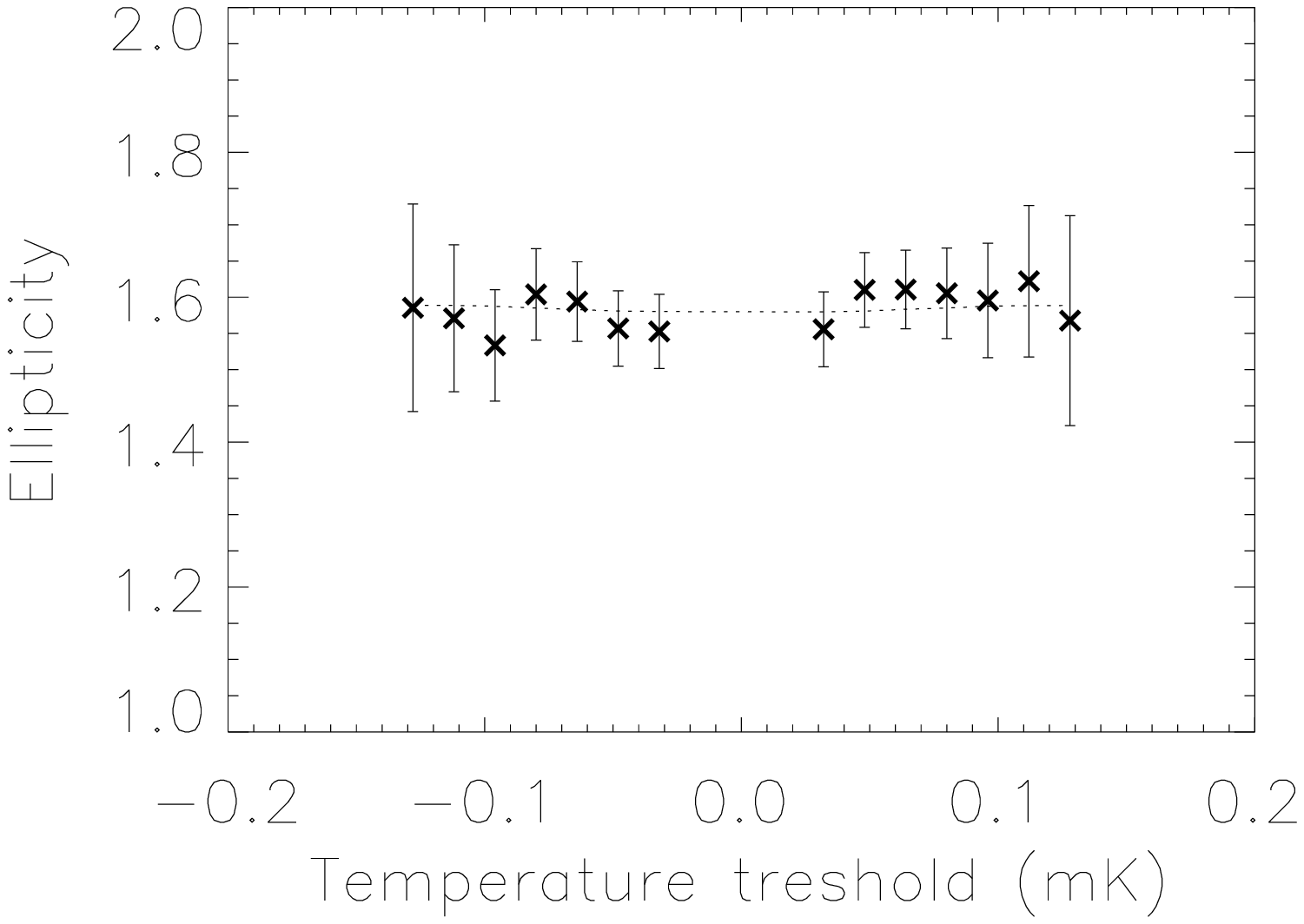,
      width=0.45\textwidth}}
  \caption{Spot ellipticity in the V+W map as a function of threshold for a given size
  of spot. Data points are shown as crosses and the mean value from simulations as a dotted line. Confidence intervals are given to $2\sigma$. Note that we only show the points for which we had sufficient spot statistics to
be used in the analysis. Note also that the points are highly correlated.}
  \label{fig:VWellip}
\end{figure}

Figure \ref{fig:colormap} shows a map of the
spots found in the case of the V+W map with a $+100 \mu \textrm{K}$
threshold. The spots have been color coded to highlight
their ellipticity in the range from 1 to 4 with zero (dark blue)
representing non--relevant pixels.
\begin{figure}[ptb]
  \centering
  \subfigure%[Full sky color coded spot map]
{\psfig{figure=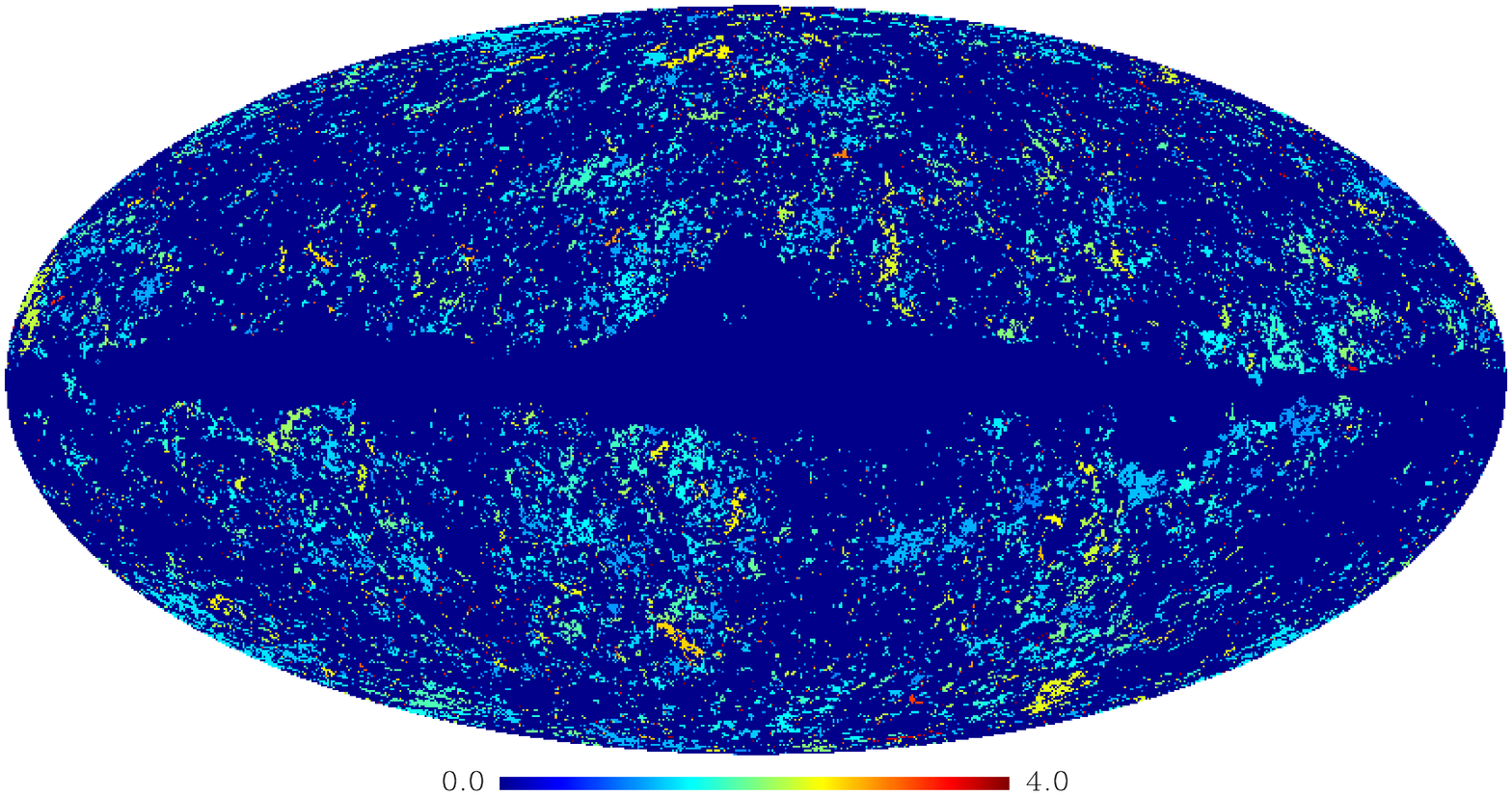, angle=90,
    width=1.0\textwidth}} 

  \subfigure%[Gnomview projection of field centered on $\phi= \frac{7
            %\pi}{9}$ and $\theta = \frac{\pi}{3}$]
  {\psfig{figure=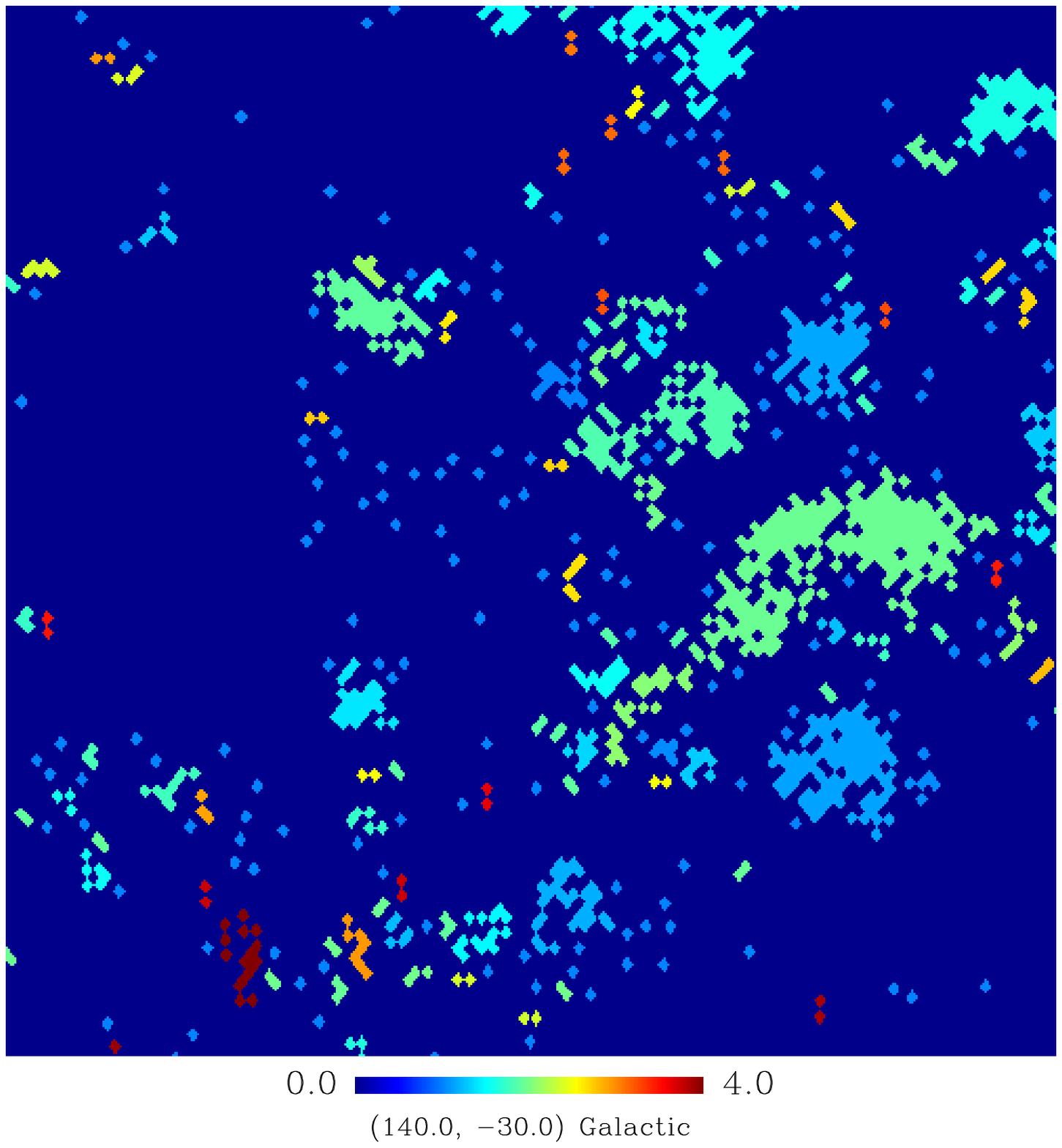,
      width=0.5\textwidth}}

  \caption{Color coded spot map for the V+W map with positive cut of
    100$\mu$K. Non relevant pixels have the value 0 and relevant
    pixels have value equal to their spots ellipticity. The zoom shows clearly some of the
    smaller spots having ellipticity of 4.0 or higher. Notice how two-pixel spots can have quite differing ellipticity depending on
    position as pixels are stretched to maintain equal area.}
  \label{fig:colormap}
\end{figure}

The measured ellipticity and obliquity was compared to an ensemble of
5000 simulated Gaussian maps and values compared to its $\chi^2$
distribution, results for which are shown in table \ref{tab:chiVW} (VW data)
and table \ref{tab:chiQ} (Q data). In table \ref{tab:totalchi} we show the total $\chi^2$ summed also over all pixel
numbers for ellipticity and obliquity as well as the skewness and kurtosis of these.
 No significant detection of
excess  ellipticity could be found for any of the tested
bands. In addition to the obliquity measured for the galactic equator, we also measured obliquity against a set of different equatorial rings orthogonal to vectors
pointing  at the center of all pixels on a $\ns=32$ map, but still no particular direction or obliquity was found.

\begin{deluxetable}{c|ccc}
%\tabletypesize{\tiny}
\tablecaption{VW band $\chi^2$ expressed as percentage of simulated maps with
  \emph{greater} $\chi^2$ than the corresponding \emph{WMAP} values. \label{tab:chiVW}} 
\tablewidth{0pt}
\tablehead{
Spot size &Number of spots&Ellipticity&Obliquity}
\startdata
%\hline
$>3$   & 74 & 20 & 15 \\
$>8$   & 73 & 13 & 29 \\
$>20$  & 43 & 82 & 42 \\
$>50$  & 20 & 10 & 79 \\
$>100$ & 67 & 22  & 44 \\
$>300$ & 25 & 81 & 90 \\
\enddata
\end{deluxetable}

\begin{deluxetable}{c|ccc}
%\tabletypesize{\tiny}
\tablecaption{Q band $\chi^2$ expressed as percentage of simulated maps with
  \emph{greater} $\chi^2$ than the corresponding \emph{WMAP} values.\label{tab:chiQ}} 
\tablewidth{0pt}
\tablehead{
Spot size &Number of spots&Ellipticity&Obliquity}
\startdata
%\hline
$>3$   & 45 & 14 & 1 \\
$>8$   & 76 & 10 & 35 \\
$>20$  & 92 & 82 & 32 \\
$>50$  & 42 & 96 & 99 \\
$>100$ & 53 & 82 & 38 \\
$>300$ & 72 & 81 & 3 \\
\enddata
\end{deluxetable}

\begin{deluxetable}{c|cc}
%\tabletypesize{\tiny}
\tablecaption{total $\chi^2$ summed over temperature threshold and pixel numbers expressed as percentage of simulated maps with
  \emph{greater} $\chi^2$ than the corresponding \emph{WMAP} values.\label{tab:totalchi}} 
\tablewidth{0pt}
\tablehead{
Spot size &VW-band&Q-band}
\startdata
%\hline
Number of spots      & 81 & 98  \\
Ellipticity          & 6 & 27  \\
Obliquity            & 13 & 2 \\
Skewness ellipticity & 46 & 1 \\
Kurtosis ellipticity & 13 & 49 \\
Skewness obliquity   & 62 & 21  \\
Kurtosis obliquity   & 60 & 58  \\
\enddata
\end{deluxetable}

\section{Discussion}
The spots in this paper were defined in the same manner as
\cite{Gurzadyan:2004mn}, as was the formula for
calculating ellipticity
\begin{equation}
\epsilon = \frac{a}{b}
\end{equation}
for a and b as major and minor axis respectively. The definition of
semi-major axis, however, was slightly different to the one presented in
\cite{Gurzadyan:2002cw}, where the semi-major axis is found by
defining the spot center and letting the semi-major axis be the line to the
center of the spot from the pixel furthest away from that center.
This difference in calculation of semi-major axis should disappear when averaged over many
spots and for sufficiently large spots the ellipticity should be close to the theoretical mean value of $\varepsilon \approx 1.648$ \citep{Aurich} in both cases. We have shown above that this is indeed the case for our algorithm, whereas in \cite{Gurzadyan:2006uk} (Figure 7)
the mean ellipticity for simulated maps is always above 2 for
any pixel size. Note that our results are based on the 7-year release of the WMAP data, whereas
\cite{Gurzadyan:2006uk} used the 3-year release. Small differences in noise
fluctuations between the two data releases will cause tiny differences in the
absolute numbers for ellipticity, but this is unlikely to be the reason for the 10 sigma 
detection of excess ellipticity in \cite{Gurzadyan:2006uk}.

Obliquity was also calculated slightly differently,
but here our results agree with \cite{Gurzadyan:2004mn} in that no
significant detection could be found.

Residual foregrounds were not considered, and confidence was placed on the
mask. The mask used (KQ85) is fairly liberal, and this may create
problems especially in the Q band; if so, the effect is small enough
not to be detectable.

In table \ref{tab:chiVW} and \ref{tab:chiQ} we presented
$\chi^2$ for ellipticity and obliquity for different spot sizes.
In these table there are no indications for excess ellipticity or obliquity for
any spot size. In order to check whether such
numbers are expected given the number of data points, we also calculated
the total $\chi^2$ taking into account all spot sizes. The results
are shown in table \ref{tab:totalchi} where we also show $\chi^2$ values
for the skewness and kurtosis of ellipticity and obliquity. 
All values seem to indicate that the ellipticity and obliquity of
spots in the WMAP data are consistent with simulated data sets based on the
concordance cosmological model.

\section{Conclusions}
\label{sect:conclusions}
Gurzadyan et al have, in several papers, claimed strong
evidence for an abnormally high ellipticity in the hot and cold spots of CMB fluctuations as measured
both by BOOMERanG and WMAP and to a certain extent also COBE
\citep{Gurzadyan:2002cw,Gurzadyan:2004mn,Gurzadyan:2005uw}. Here, the WMAP seven year data was examined in the Q,
and combined V+W band maps to look for such an effect. No
extraordinary ellipticity was found, and the results obtained here
also disagree with the reported substantial difference in
ellipticity for spots greater than 50 pixels compared to results when spots of
20 to 50 pixels also are included.

\cite{Gurzadyan:2002cw,Gurzadyan:2004mn,Gurzadyan:2005uw} also reports that no preferred direction can be
found to any statistical significance on the CMB spots. Our results
agree that no such direction can be found to
within a satisfactory statistical confidence.

Gurzadyan et al interpret the claimed ellipticity as evidence for
geodesics mixing in a hyperbolic universe
(see for instance \cite{Gurzadyan:2005uw}). This is contrary to reports suggesting that
the universe is flat; first from BOOMERanG \citep{Bernardis00aflat},
then from WMAP data \citep{2007ApJS..170..377S,Hinshaw:2008kr}
(with the latter reporting $-0.0179 < \Omega_k < 0.0089$ to $95\%$
confidence) and also contrary to claims that no difference in
ellipticity should be found even with substantial revisions of
standard cosmological models \citep{Aurich}. Not detecting any abnormal
ellipticity is thus consistent with the
current cosmological standard model. This holds true for all frequency maps
considered.

As the Planck mission will be releasing its data, redoing the analysis
on higher resolution maps will determine the extent of pixel
effects and noise in the heightened ellipticity for the smallest
spots. Probing smaller scales will not only increase the statistics
overall, but make it possible to compare ellipticity of large spots to
small ones given the increased sensitivity. But with currently available WMAP data,
we are not able to reproduce the results of excess ellipticity reported in previous papers.

\begin{acknowledgments}
FKH is thankful for a grant from the Norwegian Research Council.
We acknowledge the use of the HEALPix package and
of the Legacy Archive for Microwave Background
Data Analysis (LAMBDA). Support for LAMBDA is provided by the NASA
Office of Space Science. Support for HEALPix is provided by JPL at CalTech
and the international HEALPix crew.
\end{acknowledgments}

\end{document}